\documentclass[a4paper,10pt]{article}
\usepackage{graphicx}
\usepackage{epsfig}
\usepackage[utf8]{inputenc}

\title{STAR FORMATION RATES IN NEARBY MARKARIAN GALAXIES\footnote{Published in Astrophysics, Vol. 57, No. 1, March , 2014}}
\author{V. E. Karachentseva\footnote{Main Astronomical Observatory, National Academy of Sciences of Ukraine, Ukraine; e-mail: valkarach@gmail.com},
O. V. Melnyk\footnote{Astronomical Observatory of Taras Shevchenko Kiev National University, Ukraine; e-mail: melnykol@gmail.com, 
Institut d’Astrophysique et de G\'eophysique, Universit\'e de Li\`ge, Belgium}, and I. D. Karachentsev\footnote
{Special Astrophysical Observatory, Russian Academy of Sciences, Russia; e-mail: ikar@sao.ru} }

\begin{document}

\maketitle

\begin{abstract}
The star formation rates for the 230 nearest Markarian galaxies with radial velocities $V_{LG}<$3500 km/s
have been determined from their far ultraviolet fluxes obtained with the GALEX satellite. We briefly discuss
the observed relationship between the star formation rate and other integral parameters of these galaxies:
stellar mass, hydrogen mass, morphological type, and activity index. On the average, the Markarian
galaxies have reserves of gas that are a factor of two smaller than those of galaxies in the field of the same
stellar mass and type. Despite their elevated activity, the specific rate of star formation in the Markarian
galaxies, $SFR/M_*$, does not exceed a limit of $\sim$dex(-9.4) [yr$^{-1}$].

\end{abstract}

\section{Introduction}

In 1963 B. E. Markarian published [1] a list of 41 galaxies for which a discrepancy had been observed between
their color and morphological type, in the sense that the central parts of these galaxies have a bluer light than normal
galaxies of the same Hubble type. Thus, it was proposed [1] that the emission from the nuclei of some galaxies is
non-thermal in nature and this should be expressed as an excess of ultraviolet radiation ($UV$ excess) in the central parts.
Markarian conducted a spectral survey of the northern sky and then, together with colleagues in 1965-1981 at the 40-52$"$ 
Schmidt telescope at the Byurakan Observatory using a low-dispersion prism. The result of these many years of
work was the publication of 15 lists of galaxies with ultraviolet continua, which were then compiled to the First Byurakan
Survey-Catalog of galaxies with a $UV$ continuum [2]. The observation process, the principles for selection and
classification of the objects, and the general characteristics of the Catalog are described in detail in the Catalog. In
the worldwide literature, the objects of the catalog [2] have come to be known as Markarian galaxies. A catalog of
Markarian galaxies [3] was prepared almost simultaneously with the first survey [2]. Another version of the catalog
of Markarian galaxies, supplemented with new observational data, has been published recently [4].
The results [2-4] can be summarized as follows:

1. The term “Markarian galaxies” combines galaxies of widely different morphological types -- elliptical,
lenticular, spiral, blue compact, and irregular dwarf, as well as bright HII regions in spiral and irregular galaxies. A
comparison of Markarian galaxies [2] with objects in other catalogs and lists shows that a substantial fraction of them
can be identified with compact and post-eruptive galaxies [5,6], interacting Vorontsov-Velyaminov galaxies [7,8], and
other peculiar objects.

2. Since almost all Markarian galaxies now have measured radial velocities, it is possible to compare their
position relative to known clusters, groups, and sparsely populated systems all the way to isolated galaxies. It turns
out that the Markarian galaxies lie in systems with different multiplicities, while no more than 2\% of them are isolated galaxies.

3. Markarian galaxies manifest different degrees of nuclear activity. Their spectra include signs of quasars
(QSO), Seyfert galaxies (Sy) of types 1, 2, and intermediate types, or so called “Wolf-Rayet” (WR) galaxies, as well
as galaxies with starburst activity or with spectra similar to the spectra of HII regions. Some of the galaxies are
characterized by ordinary emission spectra (e) and a very few have absorption spectra (a). Note that the existing
diagnostic diagrams make it possible to assign galaxies to one or another activity class in more than one way.

The features of Markarian galaxies listed above explain their value for solving various problems related to the
origin and internal evolution of galaxies, as well as the influence of an interaction on active processes in galaxies.
Observational data on a large number of Markarian galaxies can be found in the citations placed in the catalogs
[2-4], as well as in conference proceedings [9] and other papers. The modern surveys make it possible to analyze
different properties of Markarian galaxies in bulk. We have made use [10] the far ultraviolet ({\em FUV}) fluxes obtained
with the GALEX satellite [11] to determine the star formation rate in nearby isolated galaxies from the LOG catalog
[12]. Here we examine the star formation characteristics of Markarian galaxies contained within the same volume of
the Local Supercluster and its surroundings as the LOG galaxies and compare them.

\section{The sample of nearby Markarian galaxies}

We have selected galaxies from the catalog [4] with radial velocities reduced to the centroid of the Local
Group, $V_{LG} <$ 3500 km/s, based on the standard used in the NED data base (http://nedwww.ipac.caltech.edu) according
to [13]. For these 260 galaxies we have calculated the distances $D = V_{LG} /H_{0}$ in Mpc for a value of $H_{0}$ = 72 km/s/Mpc. We have also used the activity class characteristics from [4]. According to the identifications,
half the chosen nearby Markarian galaxies appear in the catalogs UM, KUG, SBS, etc. (see the corresponding references
in the NED data base).

We have determined the types of Markarian galaxies (or of “parent” galaxies, part of which are Markarian
galaxies) based on their images in the Sloan Digital Sky Survey (http://sdss.eso.org) or in the Second Palomar Survey
(http://dss.eso.org) in the de Vaucoulers digital scale. Wherever possible, for a more precise classification spectra of
the galaxies from the Sloan survey were used. We noticed an inconsistency between form and spectrum in roughly
a quarter of the galaxies. We checked each of the 260 Markarian galaxies in the NED data base for the presence of
an ultraviolet flux {\em FUV} ($\lambda_{eff}$ = 1539 $\mathring{A}$, FWHM = 269 $\mathring{A}$) based on measurements by the GALEX orbital telescope
[11]. Ultraviolet fluxes were observed in 230 of the 260 objects; the absence of UV fluxes was mostly noted in
galaxies of early types. When the {\em FUV} image of a galaxy was resolved into several knots, we summed the
flux $F_{FUV}$ over the entire optical disk of the galaxy. The global rate of star formation in a galaxy, {\em SFR}, was defined
in accordance with the scheme of Lee, et al [14]:

\begin{equation}
\log(SFR[M_{\odot} yr^{-1}])=\log F^{c}_{FUV}+2\log D-6.78,
\end{equation}

where D is the distance to the galaxy (Mpc) and the flux $F_{FUV}$ (mJy) is corrected for optical extinction by

\begin{equation}
\log(F^{c}_{FUV}/F_{FUV})=0.772(A^{G}_{B}+A^{i}_{B}).
\end{equation}

Here the magnitude of the $B$-band Galactic extinction, $A^{G}_{B}$, was taken from [15] and the internal extinction, $A^{i}_{B}$, was given by
\begin{equation}
A^{G}_{B}=[1.54+2.54(\log2V_{rot}-2.5)]\log(a/b),
\end{equation}

where $a/b$ is the apparent axis ratio of a galaxy taken from NED or our own measurements and $V_{rot}$ is the
maximum corrected rotation velocity from the HyperLeda data base (http://leda.univ-lyon1.fr). For dwarf galaxies with
$V_{rot} <$ 39 km/s and gas-poor $E$ and $S0$ galaxies, the internal extinction was assumed to be zero.
The specific star formation rate is given by $SSFR = SFR/M_*$. The value of $SFR$ (in units of solar mass per
year) was determined using Eq. (1) with Eqs. (2) and (3), while the stellar mass $M_*$ of the galaxy (in units of solar
mass) was determined from its integral magnitude $K_s$ under the assumption that the ratio $M_*/L_{K}$ = 1 [17] with
$M_{K,\odot}$ = 3.28 [18]. Since most Markarian galaxies are blue objects (for which the 2MASS survey underestimates the
integral infrared magnitudes strongly) we determined the $K_s$ magnitude from the B magnitude taken from
the HyperLeda data base and the average color index: ($B - K$) = 4.10 for types $T <$ 3; ($B - K$ ) = 4.60 - 0.2$T$ for $T$ = 3 - 8; and,
($B - K$) = 2.35 for $T$ = 9 and 10 and as in [10]. The correction to the $K_s$ magnitude for Galactic and internal extinction was
defined as 0.085($A^{G}_{B}+A^{i}_{B}$).

The effective star formation rate is the star formation rate normalized to unit hydrogen mass of a galaxy, i.e.,
$SFR/M_{HI}$. The hydrogen mass (in solar units) was calculated as $M_{HI} = 2.36\cdot10^{5}D^{2}F_{HI}$, while we used the corrected
“hydrogen” magnitude $m^c_{21}$ from the HyperLeda data base to determine the flux $F_{HI}$ (Jy km/s).

The dimensionless parameters "$P$ = past" and "$F$ = future" were introduced in [19,20] to characterize
the evolutionary state of a galaxy, with

\begin{equation}
P=\log(SFR\cdot T_0 /L_{K}),
\end{equation}

\begin{equation}
F=\log(1.85\cdot M_{HI}/SFR\cdot T_0),
\end{equation}

where $T_{0} = 13.7 \cdot 10^9$ yr is the age of the universe and the coefficient 1.85 accounts for the contribution of helium and
molecular hydrogen to the overall mass of the gas [21].

\section{Some integral characteristics of the Markarian galaxies}

\begin{figure*}
\includegraphics[width=12cm,trim=15 15 60 50,clip]{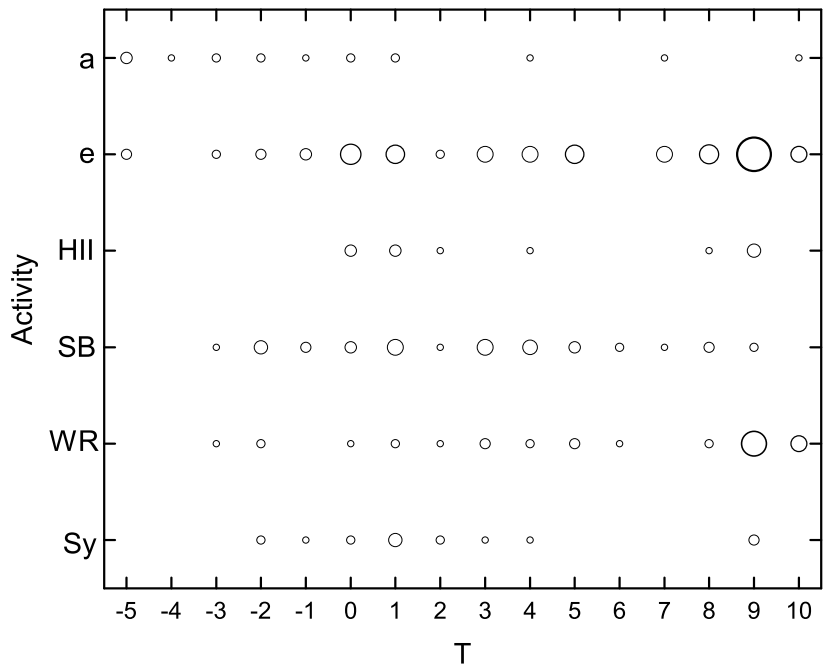}
\caption{The distribution of the 260 Markarian
galaxies with respect to morphological type and
activity index. The area of a circle is proportional
to the number of objects in the corresponding cell.}
\label{Fig1}
\end{figure*}

In the following we use the parameters of the nearby Markarian galaxies that we have determined and calculated
using the equations of Section 2 to plot the various distributions and dependences which characterize the sample as
a whole. We note that in some cases the number of galaxies is less than 260 because of the presence of one or another
value in the data bases that were used.

Figure 1 shows the distribution of the galaxies with respect to their morphological types and activity indices.
According to the de Vaucouleurs scale the elliptical galaxies have $T <$ 0, while $T (S0) =$ 0, $T (Sa)$ = 1,
$T (Sc)$ = 5, $T (Sdm)$ = 8, and $T$ = 9, 10 for the blue compact (BCD) and irregular Im and Ir dwarfs. The activity
indices are taken from [4]. No significant correlation between morphological type and activity index is observed
(the correlation coefficient $R$ = - 0.06); this can be explained by the very short time phase of the activity compared
to the time necessary to form the global structure of a galaxy.

\begin{figure*}
\begin{tabular}{c}
\epsfig{file=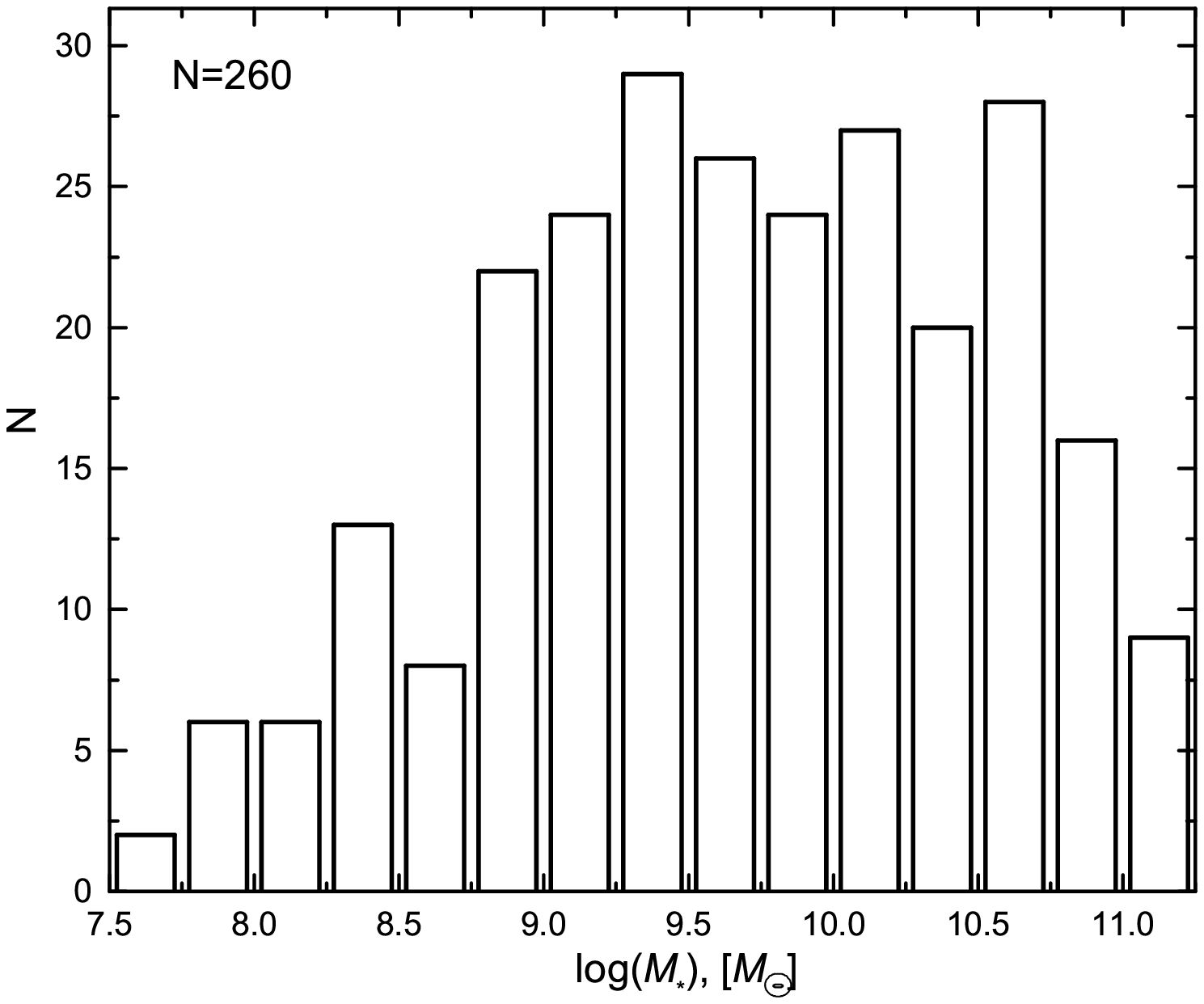,trim=15 15 40 40,clip,width=9cm} \\
\epsfig{file=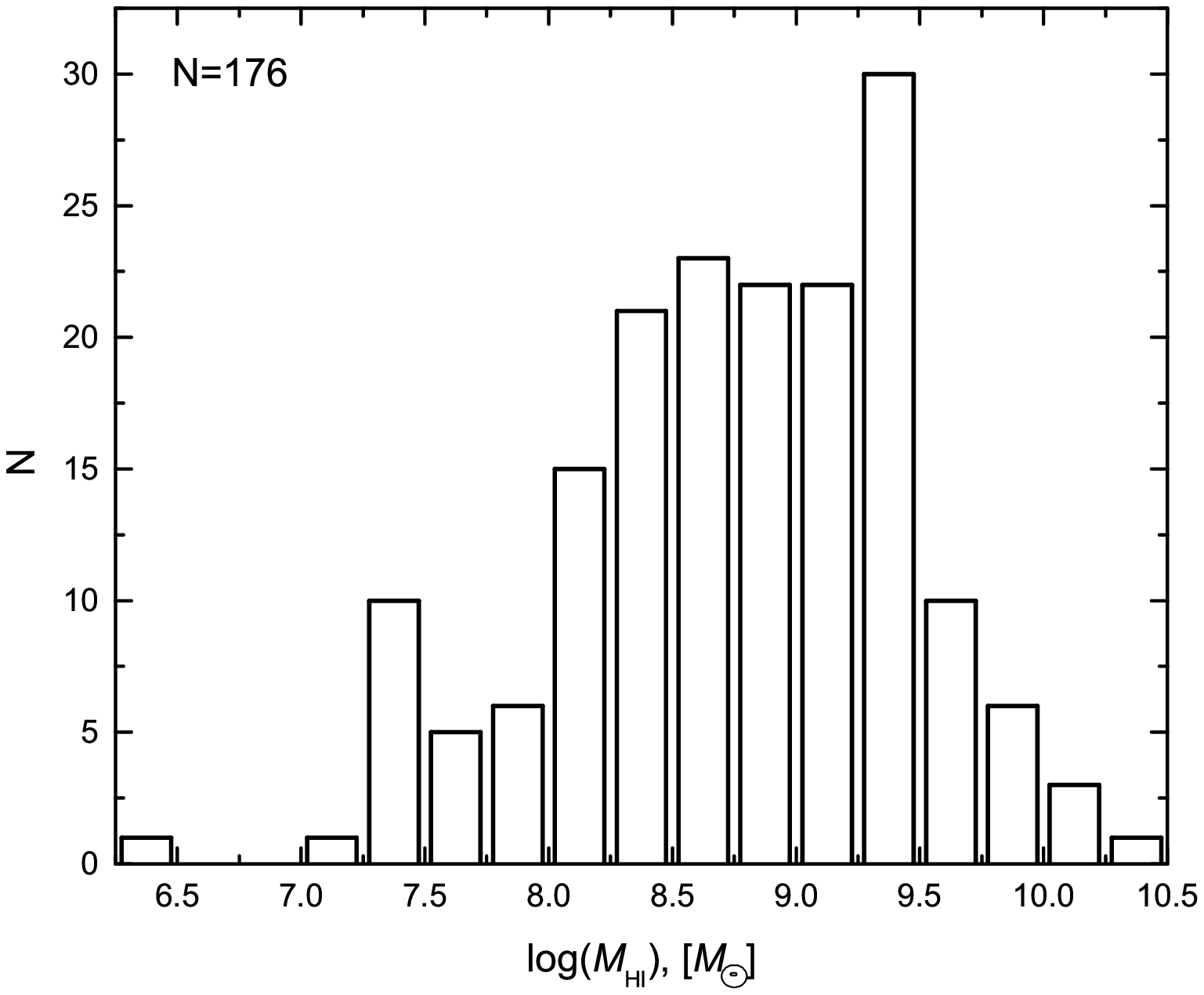,trim=15 15 40 40,clip,width=9cm} \\
\epsfig{file=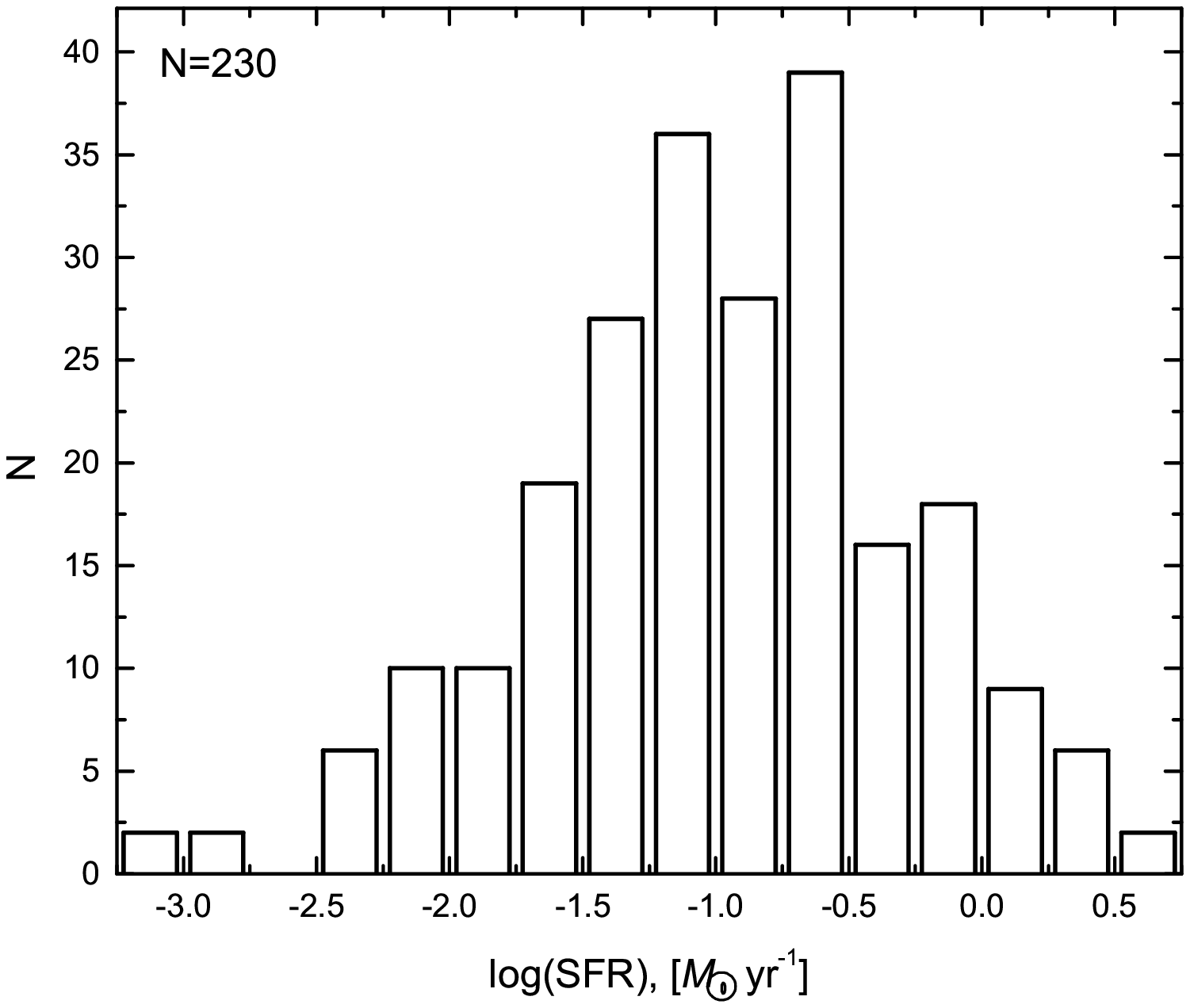,trim=15 15 40 40,clip,width=9cm} \\
  \end{tabular}
\caption{Distributions of the
Markarian galaxies with respect to
various characteristics: (a) logarithm
of stellar mass, (b) logarithm of
hydrogen mass, and (c) logarithm of
star formation rate. Here and in the
following the number of galaxies is
indicated on the graphs.}
\label{2}
\end{figure*}

The histograms of Fig. 2 show the distributions of the numbers of galaxies with respect to stellar mass (a),
hydrogen mass (b), and star formation rate (c). The corresponding medians are $med$log($M_*)$ = 9.696,
$med$log($M_{HI})$ = 8.818, and $med$log($SFR)$ = -0.969. A comparison with the data for isolated galaxies in the LOG catalog
(which are placed in the same volume) shows that the star formation rate in the Markarian galaxies is somewhat higher
than in the isolated galaxies ($med$log($SFR)$ = -1.05), with the Markarian galaxies having roughly 2.2 times as much mass
but containing less neutral hydrogen. Figure 3 shows that the Markarian galaxies manifest the well known increase
in the fraction of neutral hydrogen on going from massive (bright) to dwarf galaxies. The straight line corresponds
to the fit  $\log(M_{HI}/M_*)=-0.43\log(M_*$)+3.25 with a correlation coefficient $R$ = - 0.63 and a standard deviation $SD$
= 0.46. In this and the following figures we group the galaxies and denote these groups by the following symbols:
galaxies of early types (from -5 to 1) by solid circles, normal spirals (types from 2 to 8) by crosses, and dwarf galaxies
(9 and 10) by hollow circles. Note that the slope and spread in this plot are roughly the same as found [10] for the
isolated LOG galaxies.

\begin{figure*}
\includegraphics[width=12cm,trim=15 15 60 50,clip]{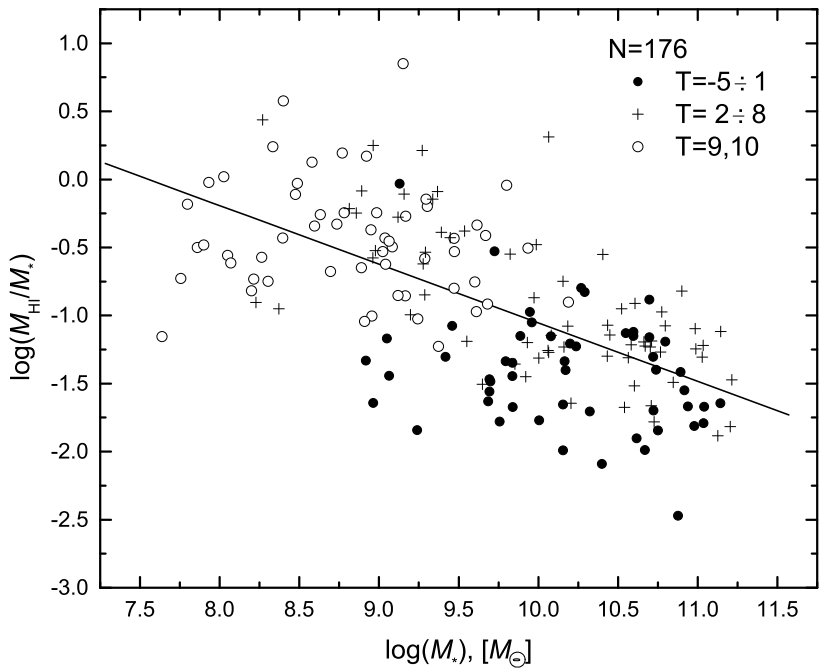}
\caption{The fraction of neutral hydrogen as a
function of stellar mass. The straight line is a
linear fit log($M_{HI}/M_{*}) = -0.43\log(M_{*}) + 3.25$ with
a correlation coefficient $R$ = -0.63 and a standard
deviation $SD$ = 0.46. The symbols for the types
of galaxies are shown at the upper right.}
\label{Fig3}
\end{figure*}

\begin{figure*}
\includegraphics[width=12cm,trim=15 15 60 50,clip]{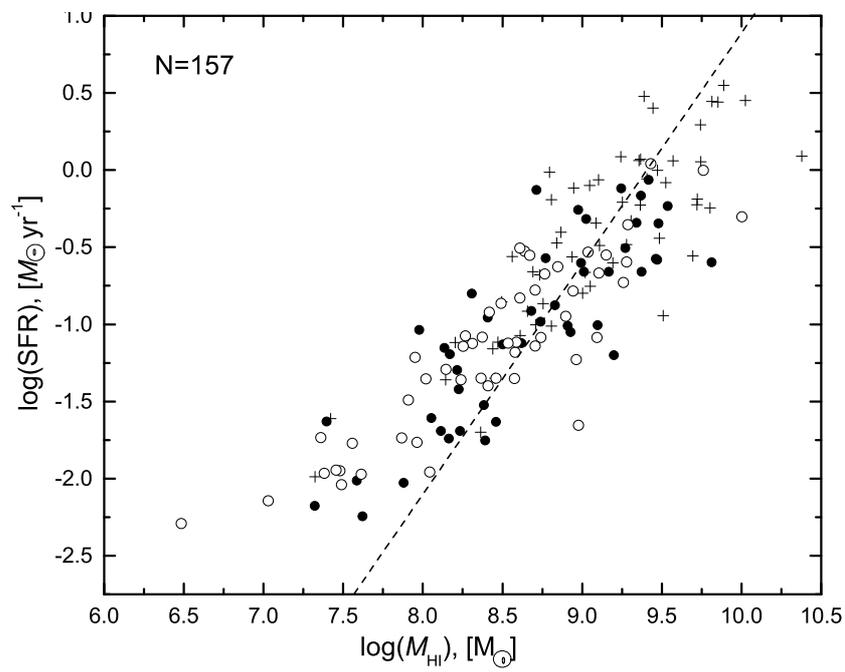}
\caption{Star formation rate as a function of
hydrogen mass. The dashed line has a slope of
3/2. The notation for the types of galaxies is the
same as in Fig. 3.}
\label{Fig4}
\end{figure*}

The Schmidt law, which has been found empirically and confirmed by many observations [22], relates the effective star 
formation rate observed in the galaxies and the surface density of gas in a power law dependence with
an exponent of roughly 1.5. The integral rate of star formation and the hydrogen mass of the galaxies as a whole also
follow this kind of dependence [23,10]. Figure 4 shows the relationship between the star formation rate and the neutral
hydrogen mass in the Markarian galaxies. The notation for the galactic types is the same as in Fig. 3. The fit line
has a slope of 1.5. It is clear that the Schmidt-Kennicutt law is obeyed satisfactorily by most of the massive galaxies.
However, for low hydrogen masses the galaxies deviate to the left from the main line. They are characterized by small
(only a few km/s) widths of the HI line and mostly belong to types 9 and 10. The increased spread in the Schmidt-
Kennicutt diagram for galaxies weaker than $M_{B}\sim$-15 and with $V_{max}\sim$50 km/s has been noticed by many authors [24],
but there is no unambiguous explanation for this fact. We note, however, that the analogous Schmidt-Kennicutt diagram for
isolated dwarf galaxies from the LOG catalog is completely symmetric (see Fig. 3 in [10]).

\section{Specific and effective star formation rates}

Figure 5a is a plot of log($SSFR)$ as a function of the absolute corrected $B$-magnitude for the galaxies grouped
according to type. The galaxies with bulges ($T$ from -5 to 1), spirals ($T$ from 2 to 8), and BCD and irregular galaxies ($T$ = 9, 10) 
distinguish clearly in this figure. The grouped types are indicated on the graph of Fig. 5a by the same
symbols as in the earlier figures. The highest specific star formation rate is observed in the faint galaxies of types
9 and 10. The spiral galaxies occupy an intermediate position in the graph and the specific star formation rate is lowest
for the galaxies with bulges. The upper limit corresponding to log($SSFR)$ = -9.4 for nearby isolated galaxies from
the LOG catalog [10] and the galaxies from the nearby volume from [23] is indicated by a horizontal line. Only
one galaxy, UGC 4499, within which Mrk 94 is identified with one of the knots in the spiral structure, lies
above it. Another galaxy with log($SSFR)$ = - 9.42 lies near this limit. It is Mrk 116 = IZw 18, a known BCD object
with record low metallicity that has been classified by Izotov, et al. [25], as a WR galaxy. The dependence of
log($SSFR)$ on the stellar mass of the galaxies (Fig. 5b) illustrates the differences in the specific star formation
rates for galaxies with high and low stellar masses more clearly than Fig. 5a. Table 1 lists the averages and standard
deviations of log($SSFR)$ for the three subgroups of morphological types.

Separation by activity class and an attempt to find a significant difference among these classes in terms of
specific star formation rate yielded no definitive result (see Table 2). Here only the sparse class of massive galaxies
with absorption spectra stands out clearly: they have an average value of log($SSFR)$ = -11.99, the lowest among all
the galaxies.

\begin{figure*}
\begin{tabular}{c}
\epsfig{file=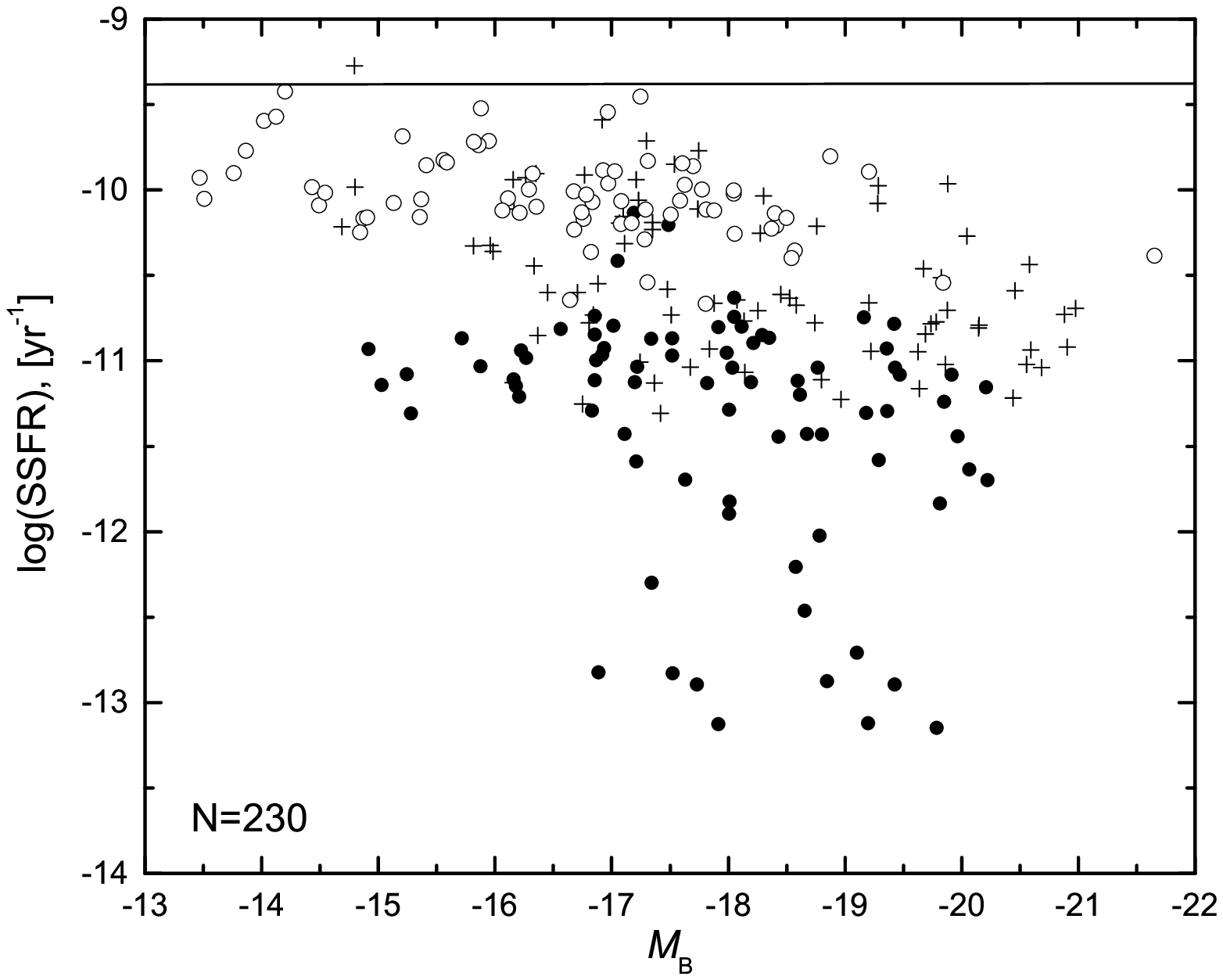,trim=15 15 40 40,clip,width=9cm} \\
\epsfig{file=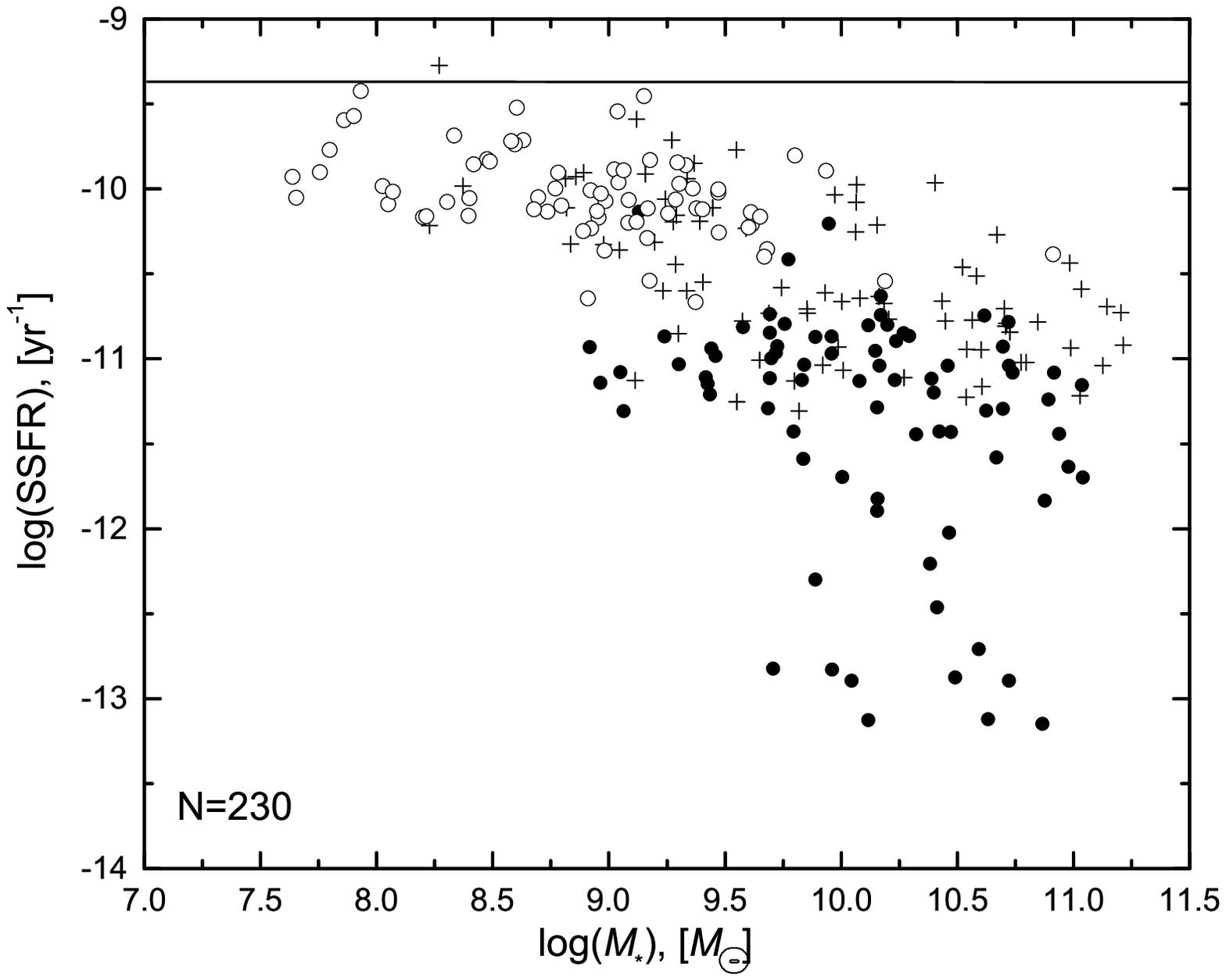,trim=15 15 40 40,clip,width=9cm} \\
\epsfig{file=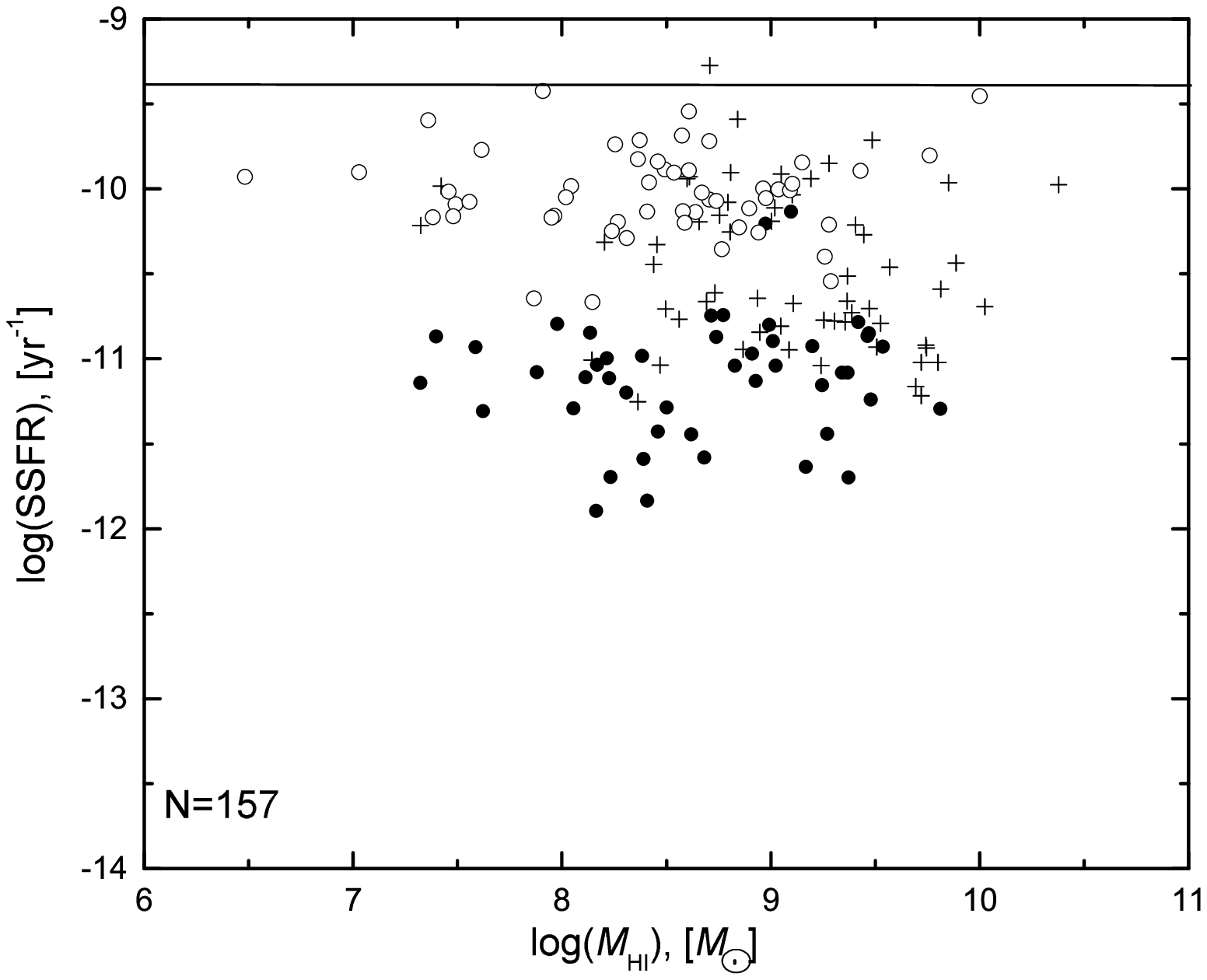,trim=15 15 40 40,clip,width=9cm} \\
  \end{tabular}
\caption{Specific star formation rate as
a function of (a) absolute
B-magnitude, (b) stellar mass of the
galaxy, and (c) hydrogen mass. The
notation for the types of galaxies is the
same as in Fig. 3.}
\label{5}
\end{figure*}

\begin{table*}
 \caption{Specific Star Formation Rate log($SSFR$) of
the Markarian Galaxies of Different Morphological Types}
 \begin{tabular}{lccc}
  \hline
  Types	& -5$\div$+1 & 2$\div$8 & 9,10 \\
  \hline
  Number & 79 & 79 & 72 \\
  Mean & -11.36 & -10.52 & -10.02 \\
  SD & 0.70 & 0.45 & 0.26 \\
  \hline
 \end{tabular}

\end{table*}

\begin{table*}
 \caption{Specific Star Formation Rate log($SSFR$) of the Markarian Galaxies
in Different Activity Classes}
 \begin{tabular}{lcccccc}
  \hline
  & WR	& HII & $e$ & Starburst & Sy & a \\
  \hline
  Number & 41 & 15 & 104 & 40 & 15 & 15 \\
  Median & -10.1 & -10.8 & -10.26 & -10.93 & -11.24 & -12.71 \\
  Mean & -10.24 & -10.61 & -10.45 & -10.89 & -11.26 & -11.99 \\
  SD & 0.56 & 0.54 & 0.55 & 0.37 & 0.81 & 1.18 \\
  \hline
 \end{tabular}

\end{table*}

\begin{table*}
\caption{Statistics of $P$ and $F$ for the Markarian Galaxies of Different Morphological Types}
\begin{tabular}{l|cccc|cccc}
 \hline
  & \multicolumn{4}{c}{Past} &  \multicolumn{4}{c}{Future} \\
Type & Median	& Mean & SD & Number & Median	& Mean & SD & Number \\
 \hline
-5$\div$+1 & -0.98 & -1.22 & 0.70 & 79 & -0.18 & -0.20 & 0.35 & 46 \\
  2$\div$8 & -0.47 & -0.38 & 0.45 & 79 & -0.37 & -0.34 & 0.34 & 58 \\
 9,10 & 0.09 & 0.11 & 0.26 & 72 & -0.31 & -0.29 & 0.33 & 53 \\
   \hline
\end{tabular}

\end{table*}

\begin{table*}
 \caption{Statistics of $P$ and $F$ for the Markarian Galaxies of Different Activity Classes}
\begin{tabular}{l|cccc|cccc}
 \hline
  & \multicolumn{4}{c}{Past} &  \multicolumn{4}{c}{Future} \\
  Activity & Median	& Mean & SD & Number & Median	& Mean & SD & Number \\
 \hline
  WR & 0.03 & -0.10 & 0.56 & 41 & -0.40 & -0.39 & 0.31 & 34 \\
 e & -0.12 & -0.32 & 0.55 & 104 & -0.28 & -0.25 & 0.36 & 63 \\
  HII & -0.66 & -0.47 & 0.54 & 15 & -0.09 & -0.20 & 0.29 & 14 \\
  Starburst & -0.75 & 0.37 & 0.26 & 40 & -0.32 & -0.20 & 0.34 & 34 \\
  Sy & -1.10 & -1.11 & 0.81 & 15 & -0.30 & -0.20 & 0.39 & 9 \\
  a & -2.56 & -1.84 & 1.18 & 15 & -0.07 & -0.09 & 0.19 & 3 \\
   \hline
 \end{tabular}

\end{table*}

\begin{figure}
\includegraphics[width=12cm,trim=15 15 60 50,clip]{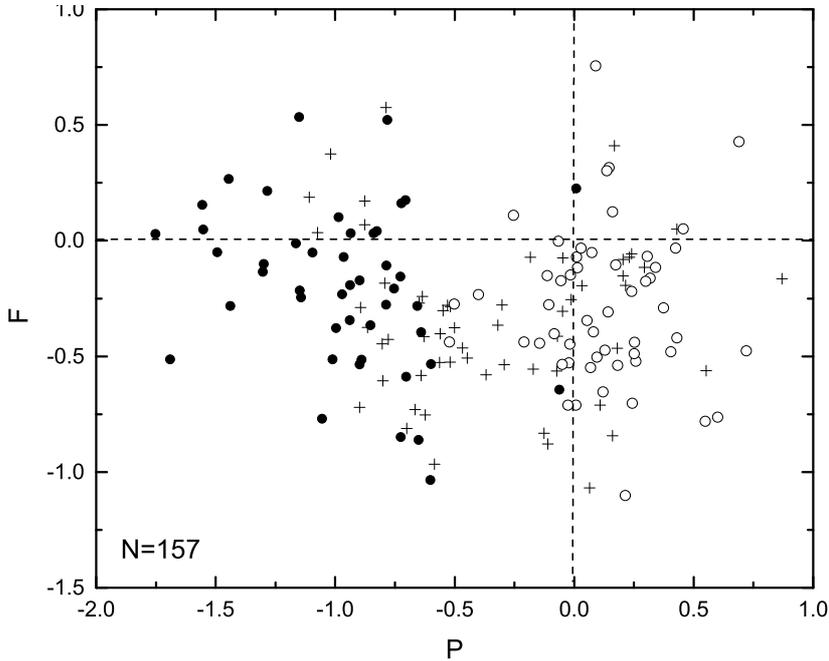}
\caption{Past-Future diagnostic diagram for the
Markarian galaxies of different types. The types
are labelled as in Fig. 3.}
\label{Fig6}
\end{figure}

\begin{figure}
\includegraphics[width=12cm,trim=15 15 60 50,clip]{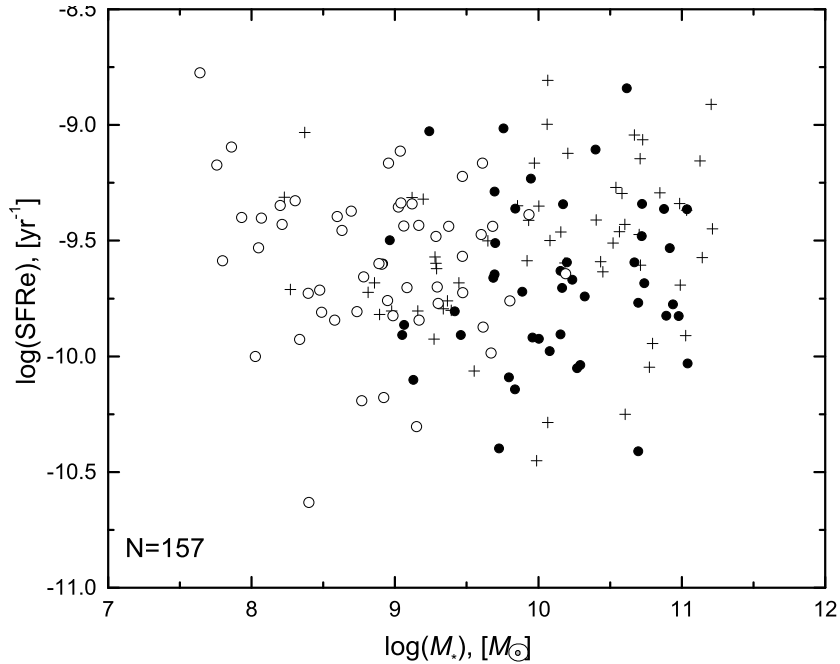}
\caption{Effective star formation rate as a function
of stellar mass of the galaxies. The notation for
the types of galaxies is as in Fig. 3.}
\label{Fig7}
\end{figure}

A clear segregation of the galaxies from early, intermediate, and late types in terms of their specific star
formation rates also shows up in the plot of log($SSFR$) as a function of hydrogen mass shown in Fig. 5c. We note
that in the case of the effective star formation rate the average and the median values of log(SFRe) do not differ
significantly either for the different types or for the different activity classes.

It has been proposed [19,20] that the evolutionary status of a galaxy be characterized by the dimensionless
parameters $P$ (past) and $F$ (future), which are independent of the errors in determining the distance to a galaxy. Figure
6 is a diagnostic ($P-F$) diagram for the Markarian galaxies separated into morphological classes. The corresponding
medians, averages, and standard deviations are listed in Table 3.

We list similar data for the different activity classes in Table 4 with ranking in decreasing order of $P$.
In terms of $P$ and $F$, a galaxy with $P$ = 0 and $F$ = 0 is capable to reproduce its observed stellar mass over
a time $T_0$ at the current star formation rate: here the reserves of gas in it are sufficient to support the observed rate
of star formation on the scale of yet another Hubble time $T_0$.

As the diagram of Fig. 6 and the data in Table 3 imply, in irregular and BCD Markarian galaxies the observed
average star formation rate is only slightly higher than the average rate over the entire cosmic scale $T_0$. However, the
reserves of gas in them will, on the average, be exhausted over roughly half the Hubble time.

In the spiral Markarian galaxies, the average current star formation rate is capable to reproduce only
$\sim$(1/2-1/3) of their stellar mass over a time $T_0$. Thus, in past epochs the star formation rate was several times the rate
being observed now. As in the case of Irr and BCD galaxies, the reserves of gas in the spiral Markarian objects will
be consumed over less time than $T_0$ on the average.

For most of the Markarian galaxies of early types ($T\leq$1), the current star formation rate is an order of
magnitude lower than the average rate in past epochs. This leads to the paradoxical conclusion that this category of
Markarian galaxies may retain its signs of high activity both during the short burst phase and over the entire cosmic
time scale $T_0$.

The data of Fig. 7 show that the effective star formation rate log($SFRe)$ remains almost constant for all values
of log($M_{*})$ from 7.5 to 11.5. It equals $\langle\log(SFRe)\rangle$=-9.68 for types -5 and +1, -9.53 for types 2 -- 8, and -9.59
for types 9 and 10. These estimates are comparable to the characteristic value log($SFRe)$ = -9.5 [yr$^-1$] obtained by
Schiminovich et al. [26], for a sample of 190 massive galaxies with log($M_*) >$ 10.0.

\section{Concluding remarks}

Numerous observations have established that star formation in galaxies in past epochs ( z $\geq$ 1 ) was an order
of magnitude more intense than in the contemporary epoch ($z <$ 0.1) [27-29]. At present the major processes for the
conversion of gas into stars take place in the disks of spiral and irregular galaxies. The distinctive feature of star
formation in disks is their protracted time scale log($\dot M_{*}$/$M_{*}$)=$SSFR^{-1} \sim 10^{10}$ yr, which is comparable to the age of the
universe $T=H_{0}^{-1}=1.37\cdot 10^{10}$ yr. The reason for the slow rate of star formation in disks is probably the existence of
a rigid feedback in this process, in which an excessively high rate of formation of young hot stars suppresses further
star formation or even entirely exhausts the reserves of neutral gas.

In an analysis of star formation in approximately 600 galaxies of the Local volume with measurements of $H_\alpha$
and FUV fluxes, Karachentsev and Kaisina [23] noted the existence of an upper bound lim(log$SSFR) = -9.4 [yr^{-1}]$ which
encompasses all the galaxies within a volume of radius 10 Mpc. Karachentsev, et al. [10], have determined the star
formation rate of 520 especially isolated galaxies in the volume of the Local supercluster of radius $\sim$50 Mpc and also
noticed the existence of this upper bound for log$SSFR$. This fact might seem trivial, since the evolution of isolated
galaxies proceeds without significant tidal influence from neighbors, which provokes star formation outbursts.
Nevertheless, as we have shown in this paper, the same upper bound on the star formation rate occurs for active
objects, i.e., Markarian galaxies. It should be noted that the large number of galaxies in the GAMA (N$\sim$70000) sample
includes objects with specific star formation rates log$SSFR\sim$-8.5 [yr$^{-1}$] [30-32]. In samples of galaxies from the
ALFALFA survey [33] and galaxies with especially low metallicity [34] it is still possible to detect galaxies with
extreme values of log$SSFR \sim [-8.0,-7.5]$. We assume, however, that these cases are artifacts arising from a large
underestimate of the stellar mass of these galaxies according to photometric data from the automatic SDSS sky survey.
In a study of the galaxies in the Local volume, Johnson, et al. [35], found no objects with a specific star formation
rate exceeding log$SSFR = - 9.2$. A similar limit was found by Gavazzi, et al. [36], for galaxies from the ALFALFA
survey in the region of the “Great Wall,” and in [37] for satellites surrounding massive galaxies of the same type as the
Milky Way. It seems obvious that verification of the cases with anomalously high estimates for the specific star
formation rate and confirmation of an upper bound on $SSFR$ will make it possible to better understand aspects of the
conversion of gas into stars. In this regard, we plan to extend the approach used in this paper to all the objects in
the Markarian catalog.

This work was supported by grants RFFI 13-02-90407-Ukr-f-a, GFFI (Ukraine) F53.2/15, and RFFI 12-02-
91338-NNIO. We have used the data bases HyperLEDA (http://leda.univ-lyon1.fr), NED (http://nedwww.ipac.caltech.edu), and SDSS (http://sdss.eso.org), 
as well as data from the Galaxy Evolution Explorer satellite (GALEX).


\begin{thebibliography}{}
\bibitem {} B. E. Markarian, Soobshch. Byurakansk. obs. {\bf 34}, 3 (1963).
\bibitem {} B. E. Markarian, V. A. Lipovetsky, J. A. Stepanian, et al., Commun. of the Special Astrophys. Obs.{\bf 62}, 5 (1989).
\bibitem {} J. M. Mazzarella and V. A. Balzano, Astrophys. J. Suppl. Ser. {\bf 62}, 751 (1986).
\bibitem {} A. Petrosian, B. McLean, R. J. Allen, and J. W. MacKenty, Astrophys. J. Suppl. Ser. {\bf 170}, 33 (2007).
\bibitem {} F. Zwicky, Catalogue of Selected Compact Galaxies and of Posteruptive Galaxies, Guemligen, Switzerland (1971).
\bibitem {} H. C. Arp, Astrophys. J. Suppl. Ser. {\bf 14}, 1 (1966).
\bibitem {} B. A. Vorontsov-Velyaminov, Atlas and Catalog of Interacting Galaxies, Sternberg Institute, Moscow (1959).
\bibitem {} B. A. Vorontsov-Velyaminov, Astron. Astrophys. Suppl. {\bf 28}, 1 (1977).
\bibitem {} “The Starburst-AGN Connection” ASP Conf. Ser. {\bf 408} (2009).
\bibitem {} I. D. Karachentsev, V. E. Karachentseva, O. V. Melnyk, and H. M. Courtois, Astrophys. Bull. {\bf 68}, 243 (2013).
\bibitem {} D. C. Martin, J. Fanson, D. Schiminovich, et al., Astrophys. J. {\bf 619}, L1-L6 (2005).
\bibitem {} I. D. Karachentsev, D. I. Makarov, V. E. Karachentseva, and O. V. Melnyk, Astrophys. Bull. {\bf 66}, 1 (2011).
\bibitem {} I. D. Karachentsev and D. I. Makarov, Astron. J. {\bf 111}, 794 (1996).
\bibitem {} J. C. Lee, A. Gil de Paz, R. C. Kennicutt, et al., Astrophys. J. Suppl. Ser. {\bf 192}, 6 (2011).
\bibitem {} D. J. Schlegel, D. P. Finkbeiner, and M. Davis, Astrophys. J. {\bf 500}, 525 (1998).
\bibitem {} M. A. W. Verheinen, Astrophys. J. {\bf 563}, 694 (2001).
\bibitem {} E. F. Bell, D. H. McIntosh, N. Katz, and M. D. Weinberg, Astrophys. J. Suppl. Ser. {\bf 149}, 289 (2003).
\bibitem {} J. Binney and M. Merrifield, Galactic astronomy, J. Binney and M. Merrifield, ed., Princeton series in astrophysics
(1998).
\bibitem {} I. D. Karachentsev and S. S. Kaisin, Astron. J. {\bf 133}, 1883 (2007).
\bibitem {} I. D. Karachentsev and S. S. Kaisin, Astron. J. {\bf 140}, 1241 (2010).
\bibitem {} M. Fukugita and P. J. E. Peebles, Astrophys. J. {\bf 616}, 643 (2004).
\bibitem {} R. C. Kennicutt, Jr., Astrophys. J. {\bf 498}, 541 (1998).
\bibitem {} I. D. Karachentsev and E. L. Kaisina, Astron. J. {\bf 146}, 46 (2013).
\bibitem {} J. C. Lee, R. C. Kennicutt, et al., Astrophys. J. {\bf 671}, L113 2007).
\bibitem {} Yu. I. Izotov, C. B. Foltz, N. G. Guseva, et al., Astrophys. J. {\bf 487}, L37-L40 (1997).
\bibitem {} D. Schiminovich, B. Catinella, G. Kaufmann, et al., Mon. Notic. Roy. Astron. Soc. {\bf 408}, 919 (2010).
\bibitem {} B. Williams, J. J. Dalcanton, L. C. Johnson, et al., Astrophys. J.  {\bf 734} L22 (2011).
\bibitem {} P. S. Behroozi, R. S. Wechsler, and C. Conroy, Astrophys. J. {\bf 770}, 57 (2013).
\bibitem {} M. D. Kistler, H. Yuksel, and A. M. Hopkins, arXiv:1305.1630 (2013).
\bibitem {} M. A. Lara-Lopez, A. M. Hopkins, A. R. Lopez-Sanchez, et al., ArXiv:1304. 3889 (2013).
\bibitem {} A. E. Bauer, A. M. Hopkins, M. Gunawardhana, et al., arXiv:1306.2424 (2013).
\bibitem {} A. S. Robotham, J. Liske, S. P. Driver, et al., arXiv:1301.7129 (2013).
\bibitem {} S. Huang, M. P. Haynes, R. Giovanelli, and J. Brinchmann, Astrophys. J. {\bf 756}, 113 (2013).
\bibitem {} M. E. Filho, B. Winkel, J. S. Almeida, et al., arXiv:1307.4899 (2013).
\bibitem {} D. D. Johnson, D. R. Weisz, J. J. Dalcanton, et al., ArXiv:1305.7243 (2013).
\bibitem {} G. Gavazzi, G. Savorgnan, M. Fossati, et al., Astron. Astrophys. {\bf 553}, A90 (2013).
\bibitem {} J. I. Phillips, C. Wheeler, M. Boylan-Kolchin, et al., arXiv:1307.3552 (2013).


\end{thebibliography}
\end{document}